\documentclass[preprintnumbers,superscriptaddress,nofootinbib,aps,prd,floatfix]{revtex4}

\usepackage{amssymb,amsmath,multirow,graphicx,tabularx,epsfig}
\usepackage{color}
\usepackage{float}
\usepackage{multirow}
\usepackage{subfigure}
\usepackage{hyperref}

\newcommand\one{\leavevmode\hbox{\small1\normalsize\kern-.33em1}}

\newcommand{\qqquad}{\qquad \qquad}
\newcommand{\qqqquad}{\qquad \qquad \qquad}

\newcommand{\gev}{{\ensuremath\rm GeV}}

\newcommand{\ifb}{{\ensuremath\rm fb^{-1}}}

\def\slashchar#1{\setbox0=\hbox{$#1$}           
   \dimen0=\wd0                                 
   \setbox1=\hbox{/} \dimen1=\wd1               
   \ifdim\dimen0>\dimen1                        
      \rlap{\hbox to \dimen0{\hfil/\hfil}}      
      #1                                        
   \else                                        
      \rlap{\hbox to \dimen1{\hfil$#1$\hfil}}   
      /                                         
   \fi}

\def\eg{{\sl e.g.} \,}
\def\ie{{\sl i.e.} \,}

\begin{document}

\title{Improving Higgs plus Jets analyses through Fox--Wolfram Moments}

\author{Catherine Bernaciak}
\affiliation{Institut f\"ur Theoretische Physik, Universit\"at Heidelberg, Germany}

\author{Bruce Mellado}
\affiliation{School of Physics, University of the Witwatersrand, Johannesburg, South Africa}

\author{Tilman Plehn}
\affiliation{Institut f\"ur Theoretische Physik, Universit\"at Heidelberg, Germany}

\author{Peter Schichtel}
\affiliation{Institut f\"ur Theoretische Physik, Universit\"at Heidelberg, Germany}

\author{Xifeng Ruan}
\affiliation{School of Physics, University of the Witwatersrand, Johannesburg, South Africa}

\begin{abstract}
It is well known that understanding the structure of jet radiation can
significantly improve Higgs analyses. Using Fox--Wolfram moments we
systematically study the geometric patterns of additional jets in weak
boson fusion Higgs production with a decay to photons. First, we find
a significant improvement with respect to the standard analysis based
on an analysis of the tagging jet correlations. In addition, we show
that replacing a jet veto by a Fox-Wolfram moment analysis of the
extra jet radiation almost doubles the signal-to-background
ratio. Finally, we show that this improvement can also be achieved
based on a modified definition of the Fox--Wolfram moments which
avoids introducing a new physical scale below the factorization
scale. This modification can reduce the impact of theory uncertainties
on the Higgs rate and couplings measurements.
\end{abstract}

\maketitle

\tableofcontents

\newpage

\section{Introduction}
\label{sec:intro}

After the recent Higgs discovery by ATLAS and
CMS~\cite{higgs,atlas,cms}, the careful and systematic study of Higgs
properties is becoming a key research program at the LHC and a future
linear collider~\cite{ilc}. The theoretical implications of the first
fundamental scalar particle include many open questions, including the
actual generation of a vacuum expectation value, the stability of its
physical mass, or the link between the Higgs potential at the weak
scale to high--scale structures~\cite{rge_studies}. In the language of 
quantum field theory we need to construct the weak--scale Higgs
Lagrangian including the operator basis and the corresponding
couplings~\cite{couplings}.

At the LHC the weak boson fusion production channel
(WBF)~\cite{wbf_tau,wbf_w,wbf_gamma,wbf_ex,review} plays an important role in
answering some of these question, in particular once the LHC runs
closer to its design energy. It allows us to directly probe the
unitarization of $WW \to WW$ scattering and carries information on
tree--level Higgs couplings with negligible impact of perturbative
extensions of the Standard Model. Experimentally, two forward tagging
jets are highly effective in reducing QCD backgrounds~\cite{tagging},
which means that Higgs analyses in weak boson fusion typically benefit
from a signal--to--background ratio around unity.\bigskip

As an analysis tool utilizing the unique QCD structure of weak boson
fusion we rely on a central jet
veto~\cite{scaling,manchester,jetveto,neubert,gavin,scet}. It is based
on the fact that we can generate large logarithms and increase central
jet radiation in QCD backgrounds while leaving the jet activity in the
signal at low level.  This shift from staircase scaling of jets (with
constant ratios between successive exclusive jet bins) in signal and
background to staircase scaling in the signal and Poisson scaling in
the background can be derived from first--principles
QCD~\cite{scaling}. The resulting jet veto survival probabilities for
the the QCD backgrounds can be measured in data.  Their calculation
from QCD is plagued with significant theory uncertainties which in
turn will soon dominate the extraction of the Higgs couplings at the
LHC~\cite{couplings}. In addition, a jet veto always removes a wealth
of kinematic information carried by these jets, so the question arises
whether the information from the jets recoiling against the Higgs
cannot be used more efficiently.\bigskip

To answer the question of how much information is encoded in the jet
activity of Higgs candidate events we need to systematically study
multi-jet kinematics. For example in flavor physics Fox--Wolfram
moments (FWM) are an established tool to analyze such geometric
patterns~\cite{fwm_orig}, but they have hardly been employed by the
ATLAS and CMS collaborations. By construction, they are particularly
well suited to study the geometry of tagging jets in weak boson
fusion~\cite{first}. Dependent on the specific construction of their
weights the moments can also be sensitive measures of the additional
jet activity in an event. Ideally, they will enhance a central jet
veto defined on a fixed phase space region to some kind of weighted
jet veto over phase space regions based on the kinematics of the hard
process. Moreover, by choosing different weights the moments can be
adjusted such that they avoid introducing a fixed scale below the
factorization scale of the hard process. At the expense of the
background rejection efficiency they can be tuned to introduce smaller
theory uncertainties. This will allow the ATLAS and CMS experiments to
optimize their Higgs analyses including theory uncertainties and
significantly improve the case for a luminosity upgrade based on Higgs
couplings measurements.\bigskip

In this paper we will attempt to answer three questions based on the
weak boson fusion analysis with a Higgs decay to photons. This
includes a study of the signal process, the Higgs background from
gluon fusion, and the continuum production of a photon pair with jets:
\begin{enumerate}
\item in Section~\ref{sec:2jets} we will apply Fox--Wolfram moments to
  the kinematics of the two tagging jets only. Based on a multivariate
  analysis we will estimate how much these additional observables can
  improve the current ATLAS results at 8~TeV collider energy.
\item in Section~\ref{sec:veto} we will compare the performance of a
  set of Fox--Wolfram moments with a specific (unit) weight in
  comparison to the usual central jet veto for the 13~TeV run.
  Moreover, a multivariate analysis of Fox-Wolfram moments allows us
  to define a ROC curve with a free choice of operating points.
\item in Section~\ref{sec:matched} we will introduce a new weight in
  the Fox--Wolfram moments. It avoids introducing a physical momentum
  scale for the jet veto which lies below the factorization
  scale.
\end{enumerate}
Obviously, our conclusions are immediately applicable to ongoing and
future LHC analyses. Fox--Wolfram moments have been tested in a few
ATLAS and CMS analyses, so it should be a simple task to also include
them in Higgs analyses.

\section{Setting the Stage}
\label{sec:basics}

The analysis presented in this paper will give an estimate of
the impact which Fox--Wolfram moments computed from jets can have on
current and future LHC Higgs analyses. Fox--Wolfram moments are one
way to systematically evaluate angular correlations between jets in
terms of spherical harmonics. While such approaches are standard for
example in cosmology, they are largely missing in LHC physics. We will
summarize their main features below. For a more detailed account of
the WBF-specific properties we refer to an earlier paper~\cite{first}.

To allow for significant correlations between different moments we 
employ multivariate methods. Our analysis will largely be based on
boosted decision trees (BDTs), which we will also briefly introduce
below. Part of the analysis we cross--check with a neural net to 
make sure our findings are independent of the MVA method used.

\subsection{Fox--Wolfram moments}
\label{sec:moments}

Most analyses of QCD jets at the LHC are based on an {\sl ad-hoc}
selection of angular correlation variables, which have been shown to
separate signals from backgrounds. For analyses where each
one--dimensional or two--dimensional distribution is carefully
understood in terms of the underlying physics and then tuned to the
best cut value, this approach is natural and appropriate. For
multivariate analyses, where events are classified in terms of a
more generic set of kinematic observables, the choice of observables
should be more systematic. 

For angular correlations, we know how to generally describe underlying
objects, in our case jets, in terms of spherical harmonics.
Obviously, Fox--Wolfram moments do not have to be based on jets. They
are closely related to event shapes~\cite{event_shapes}, and for
example at LEP they were based on calorimeter information. At the LHC,
particle flow objects or topoclusters might eventually turn out more
useful. In this analysis we use jets to avoid additional experimental
or theoretical complications, for example due to pile-up or underlying
event.

Fox--Wolfram moments are constructed by summing jet--jet correlations
over all $2\ell+1$ directions, including an unspecified weight
function $W_i^x$~\cite{fwm_orig}
\begin{alignat}{1}
H_\ell^x = \frac{4\pi}{2\ell+1}
       \sum_{m=-\ell}^\ell \;
       \left| \sum_{i=1}^N  W_i^x \; Y_\ell^m(\Omega_i) 
       \right|^2 \; .
\label{eq:fwm_def1}
\end{alignat}
The index $i$ sums over all final state jets defined by
appropriate acceptance and selection criteria.  The general coordinates
of the spherical harmonics $Y_\ell^m(\theta,\phi)$ we replace by a
reference angle $\Omega$. The moments can be rewritten as
\begin{alignat}{2}
H^x_\ell = \sum_{i,j=1}^N \; W_{ij}^x \; P_\ell(\cos \Omega_{ij}) 
\qqquad \text{with} \quad 
W_{ij}^x = W_i^xW_j^x \; .
\label{eq:fwm_def2}
\end{alignat}
The angle $\Omega_{ij}$ is the total angle between two jets. The
weight function $W_{ij}^x$ can be chosen freely. In
Sections~\ref{sec:2jets} and \ref{sec:veto} we will use
transverse--momentum and unit weights~\cite{first}:
\begin{alignat}{1}
W_{ij}^T = \frac{p_{T i}\,p_{T j}}{\left(\sum p_{T i}\right)^2}  
\qqqquad 
W_{ij}^U = \frac{1}{N^2} \; .
\label{eq:weight_1}
\end{alignat}
The advantage of the transverse--momentum weight is that soft and
collinear jets with their limited amount of information about the hard
process are automatically suppressed. The resulting analysis becomes
stable with respect to the parton shower and QCD jet radiation. For
tagging jets without an actual collinear divergence the transverse
momentum weight should be appropriate.

Whenever we are interested in the color structure of the event, this
jet radiation will carry the crucial information. For studies of
central jet radiation we therefore expect the unit weight to be the
most promising.

In analogy to a jet veto, Fox--Wolfram moments with unit weight
introduce an energy or momentum scale, above which we include jets in
the moments. Because of the unit weight there does not exist a smooth
transition regime; requiring any Fox--Wolfram moment of such
additional jets to be different from zero corresponds to a step
function in counting the number of jets.  Because the new momentum
scale usually resides below the factorization scale of the hard event,
fixed--order precision predictions are not applicable, and a dedicated
resummation is a theoretical
challenge~\cite{scaling,manchester,jetveto,neubert,gavin,scet}.  In
Section~\ref{sec:matched} we introduce the {\sl matched weight}
\begin{alignat}{1}
W_{ij}^M = \frac{(p_{T i}-p_T^\text{min}) \, (p_{T j}-p_T^\text{min})}{\left(\sum p_{T i} - p_T^\text{min} \right)^2}  
\label{eq:weight_2}
\end{alignat}
in order to reduce the theoretical uncertainty in comparing measured
cross sections to QCD predictions. This new weight avoids introducing
a new hard scale and will be less dominated by the momentum scale
$p_{T j}^\text{min}=20$~GeV, above which jets contribute to the
Fox--Wolfram moments.

\subsection{Event generation}
\label{sec:generate}

While the description of the tagging jets in weak boson fusion is
straightforward, the continuum background with its QCD jet activity is
more tricky. Moreover, the correct description of the QCD activity in
the Higgs signal requires a careful treatment of the color structure
of the hard process. Throughout this analysis we use
\textsc{Sherpa}~\cite{sherpa} with \textsc{Ckkw} merging~\cite{ckkw}.
For the weak boson fusion signal we generate samples including up to
three hard jets, including the tagging jets. Gluon fusion Higgs
production we simulate with up to three hard jets. For the QCD
background we include di--photon production plus up to two
hard jets.  For jet clustering we rely on the anti-$k_T$ algorithm
as in \textsc{Fastjet}~\cite{fastjet} with $R=0.4$.

The assumed Higgs mass value is 126~GeV. Our cuts are dominated by the
detector acceptance and jet--photon separation,
\begin{alignat}{5}
p_{T \gamma} > 14~\gev
\qqquad 
R_{\gamma j} > 0.3
\qqquad 
m_{\gamma \gamma} > 80~\gev \; .
\label{eq:photon_cuts}
\end{alignat}
After those cuts we are left with a weak--boson--fusion signal cross
section times branching ratio of 5.2~fb at 8~TeV collider energy and
9.24~fb at 13~TeV collider energy.  To allow for an efficient
generation of background events we do not require a mass window for
the two photons in the background generation. Later in the analysis we
add an $m_{\gamma\gamma}$ window of $\pm 10$~GeV around the Higgs
mass.  For a proper Higgs analysis we should require an $m_{\gamma
  \gamma}$ window of 1-2~GeV around the measured Higgs mass. However,
with this condition the event generation for the background becomes
highly inefficient.  Because our analysis does not intend to predict
the actual signal and background cross sections and instead focuses on
the improvement over the established experimental
analysis~\cite{atlas_default}, the loose cuts of
Eq.\eqref{eq:photon_cuts} allow for a much more efficient event
generation and will not affect our conclusions.

\subsection{Boosted decision trees}
\label{sec:bdt}

Any multivariate analysis is based on some kind of mapping of a set of
observables onto a single--valued quantity, the classifier
response. Based on this classifier response we define a
classification rule to separate signal and background events.
Training the multivariate analysis on a set of simulated events aims
to determine the best classification rule for a given signal and
background. The optimal classification rule has to be determined by
some measure, for example the signal efficiency, the statistical
significance, or the signal--to--background ratio. Independent of this
optimization, we can quantify the performance of any classification
rule in terms of the signal efficiency and the background
mis-identification probability. In this two--dimensional plane we can
describe cuts on the same response parameter as a receiver operating
characteristics (ROC) curve. Given such a ROC curve we are free to
choose one or more operating points.  In line with the ATLAS di-photon
analysis we use a fixed 40\% signal efficiency $\epsilon_S$ after acceptance 
with a variable background rejection $1 - \epsilon_B$
as the standard working point. In
Section~\ref{sec:veto}, we quote the main results of our BDT
analysis for the best possible significance $S/\sqrt{S+B}$ given the
set of kinematic observables and Fox--Wolfram moments.\bigskip

Decision tree algorithms --- as they are utilized in high energy
physics applications ---  are based on a set of kinematic variables,
intended to separate signal and background events.  In the first step
they choose the `root node' variable, \ie the variable with the best
separation between signal and background.  There exist several types
of separation which we can choose from in
\textsc{Tmva}~\cite{tmva}. We use the cross entropy
\begin{alignat}{2}
C_E = -\frac{S}{S+B} \log_2 \frac{S}{S+B} 
      -\frac{B}{S+B} \log_2 \frac{B}{S+B} \; ,
\label{eq:cross}
\end{alignat}
where $S$ and $B$ are the numbers of signal and background events in a
particular subset of events.  This measure is the closest to the
original definition of information entropy~\cite{C45}.  After choosing the root
node, the subsequent nodes are ordered by their separation at some
threshold value.

For the complete decision tree the events are classified as
signal--like or background--like by some measure.  In the training set
we know how good the tree is at classifying the events. Our training
set include 100000 events for each signal and background channel. In
the next step the algorithm corrects for mistakes through a
reweighting procedure, builds another decision tree, tests its
performance, and repeats for some user--defined number of iterations.
For this `boosting' procedure we mainly use the adaptive boost algorithm
implemented in \textsc{Tmva}~\cite{tmva}. The final classification
rule for signal versus background events we then apply to an
independent event sample, again including 100000 events per signal and
background process.  To prevent over--training we limit our forest to
400 trees, and the individual trees to three layers.\bigskip

Because correlations between the different Fox--Wolfram moments are a
key issue of our systematic approach to kinematic input variables, we
carefully test two different boosting algorithms (adaptive and gradient
boost~\cite{tmva}) as well as different multivariate analysis methods.
{\sl Per se}, boosted decision trees
are not particularly well suited for studying strongly correlated
variables. The reason is that trees are built out of the individual
variables. Two strongly correlated variables are best mapped through
individual fine binnings in each of them, so a careful mapping of
correlations will eventually lead to statistical limitations and a
possible training on statistical fluctuations.  Therefore, we compare
BDT results to results using a multi--layer perceptron (MLP) neural
network whenever an independent test appears sensible. We utilize a
MLP neural network with a single hidden layer containing $N+5$ neurons,
where $N$ is the number of training variables. 

\section{Tagging jet correlation}
\label{sec:2jets}

In this first analysis we are going to use Fox--Wolfram moments to
systematically test the completeness of the tagging jet correlations
included by ATLAS. Because we directly refer to the current ATLAS
result we use a collider energy of 8~TeV for the most recent LHC run.
The two $p_T$-ordered tagging jets have to fulfill either of the two conditions
\begin{alignat}{5}
p_{T j} &> 25~\gev \quad \text{for} \quad & |y_j| &< 2.4
\notag \\
p_{T j} &> 30~\gev \quad \text{for} \quad & 2.4 \leq |y_j|& < 4.5 \; .
\label{eq:jetcut1}
\end{alignat}
These two tagging jets must also pass
\begin{alignat}{2}
|\Delta y_{j_1j_2}| \geq 2  \quad  \text{ and } \quad m_{j_1j_2} >  150~\gev
\; . 
\label{eq:jetcut2}
\end{alignat}
These cuts correspond to the variables used in the multivariate di-photon
Higgs analysis by ATLAS~\cite{atlas_default},
\begin{alignat}{2}
\{ m_{j_1j_2}\, , \, y_{j_1}\, , \, y_{j_2} \, , \,\Delta y_{j_1j_2} \} 
\qqquad  \text{(ATLAS default)} . 
\label{eq:atlas_default}
\end{alignat}
\bigskip

The angular correlations between the tagging jets in
weak--boson--fusion Higgs production is known to reflect the tensor
structure of the $WWH$ vertex~\cite{wbf_spin}. In this application the
collinearity of the two tagging jets plays an important role, with the
effect that the azimuthal angle between the tagging jet is a more
sensitive probe than the opening angle between them. For the
Fox--Wolfram moments this means that the definition in terms of the
opening angle $\Omega_{ij}$ is not optimally
suited. For the tagging jet analysis we therefore replace the opening
angle in the Legendre polynomials by the azimuthal angle
$\Delta \phi_{ij}$ between the two tagging jets,
\begin{alignat}{2}
H^{x,\phi}_\ell
= \sum_{i,j=1}^2 \; W_{ij}^x \; P_\ell(\cos \Delta \phi_{ij}) 
\; .
\label{eq:fwm_def3}
\end{alignat}
For a systematic study of the usefulness of the tagging jet
correlations we perform a multi-variate analysis of the Fox--Wolfram
moments introduced in Section~\ref{sec:moments}. Because the moments are
based on spherical harmonics they form a basis and include all
available information, given the weight $W_{ij}^x$ we use in their
definition. \bigskip

\begin{table}[t]
\begin{tabular} {l|c|c|c||c|c|c}
\hline
                   & \multicolumn{3}{c||}{BDT} & \multicolumn{3}{c}{MLP}      \\ \hline
$\epsilon_S = 0.4$ & $1 - \epsilon_B$ & $\dfrac{S}{\sqrt{S+B}}$ & $\dfrac{S}{B}$     
                   & $1 - \epsilon_B$ & $\dfrac{S}{\sqrt{S+B}}$ & $\dfrac{S}{B}$  \\ \hline
ATLAS default Eq.\eqref{eq:atlas_default}  
& 0.887  & 1.50  & 0.76 & 0.888  & 1.50  & 0.78  \\\hline \hline
$H_1^{T,\phi}$ $\rightarrow$ $H_4^{T,\phi}$, $H_1^{U,\phi}$ $\rightarrow$ $H_4^{U,\phi}$ 
& 0.952  & 1.65 & 1.54  & 0.953  & 1.65 & 1.55  \\ \hline 
$H_1^{T,\phi}$, $H_3^{T,\phi}$, $H_1^{U,\phi}$, $H_3^{U,\phi}$ 
& 0.952  & 1.66 & 1.56  & 0.952  & 1.65 & 1.54  \\ \hline  
$H_1^{T,\phi}$, $H_2^{T,\phi}$, $H_2^{U,\phi}$, $H_2^{U,\phi}$              
& 0.953  & 1.65 & 1.47  & 0.953  & 1.65 & 1.55  \\ \hline
$H_1^{T,\phi}$, $H_1^{U,\phi}$              
& 0.953  & 1.65 & 1.43  & 0.952  & 1.65 & 1.46 \\ \hline 
$H_1^{T,\phi}$                              
& 0.950  & 1.63 & 1.45  & 0.950  & 1.63 & 1.44\\ \hline
$H_1^{U,\phi}$                              
& 0.952  & 1.65 & 1.40  & 0.952  & 1.65 & 1.44 \\ \hline \hline 
$\cos\Delta\phi_{12}$, $W^T_{12}$                 
& 0.952  & 1.65 & 1.53  & 0.952  & 1.65 & 1.50\\ \hline  
$\cos\Delta\phi_{12}$                             
& 0.952  & 1.65 & 1.42  & 0.952  & 1.65 & 1.44 \\  \hline  
\end{tabular}
\caption{BDT and MLP results including azimuthal--angle Fox--Wolfram moments
  based on the two tagging jets only after Eq.\eqref{eq:jetcut2}. The background rejection is
  given for 40\% signal efficiency. The value for $S/\sqrt{S+B}$ we
  compute for an integrated luminosity of $30~\ifb$.  All
  sets of variables subsequent to the first row contain the default
  variables as well.}
\label{tab:tagging_2jonly}
\end{table}

We show some sample BDT and MLP results based on the azimuthal moments in
Table~\ref{tab:tagging_2jonly}. The full set of moments for each
weight function by definition includes all available information for
the corresponding weights.  First, we see that including a large set
of Fox--Wolfram moments gives a significant improvement of the current
ATLAS set of observables, defined in Eq.\eqref{eq:atlas_default}. Both
multivariate analyses using the first four moments with unit weight as well as
with transverse--momentum weight reduces the remaining fraction of
background events by a factor two. From the \textsc{Tmva} output we
have checked that these eight moments dominate the distinctive power of
the analysis.

Obviously, the next question is which of the Fox--Wolfram moments
contribute most to this improvement. From the earlier
analysis~\cite{first} we know that lower moments will dominate in the
tagging jet analysis, and that only odd moments can distinguish
between forward--backward and forward--forward tagging
jets. Individually, we find that the six best individual moments are
(in order) $H_1^{U,\phi}$, $H_1^{T,\phi}$, $H_3^{U,\phi}$,
$H_3^{T,\phi}$, $H_2^{U,\phi}$, and $H_2^{T,\phi}$.\footnote{Given
  that \textsc{Tmva} gives an ordered list of the most relevant
  observables, it is not clear to one of the authors (TP) why this
  very interesting information is never shown in
  experimental publications.}  The moments with unit weight are
slightly more powerful than the transverse--momentum weight. The most
striking feature is that for the tagging jet the higher moments play
hardly any role in improving the analysis.

As a matter of fact, the single moment $H_1^{U,\phi}$ is, within
uncertainties due to the training procedure, almost as powerful as the
set of the first 20 moments, both with unit and transverse--momentum
weight. Given that the corresponding Legendre
polynomial is $P_1( \cos \Delta \phi_{ij} ) = \cos \Delta \phi_{ij}$
we can further simplify the analysis by separating the
transverse--momentum weight from the azimuthal angle.  Compared to the
ATLAS default variables, adding the azimuthal angle between the
tagging jets, $\Delta \phi_{ij}$, almost doubles the
signal--to--background ratio. Systematically including the
Fox--Wolfram moments increases the signal--to--background ratio additionally
by 8\%. This result persists between the two multivariate
methods and we conclude that our improvement is truly due to the nature
of the moments and not to some advantageous choice of methods and/or
parameters for our multivariate analyses.\bigskip

Following the tagging jet analysis in this section we extend the
default set of tagging jet cuts Eq.\eqref{eq:atlas_default} for the
remainder of this paper to include
\begin{alignat}{2}
\{ m_{j_1j_2}\, , \, y_{j_1}\, , \, y_{j_2} \, , \,\Delta y_{j_1j_2} , \,\Delta \phi_{j_1j_2} \} 
\qqquad  \text{(WBF default)} . 
\label{eq:wbf_default}
\end{alignat}
It could be argued that adding the azimuthal angle to the list of
kinematic variables employed in the background rejection will make the
analysis result less applicable to modified Higgs--like signal
hypotheses. Indeed, the azimuthal angle between the tagging jets is
the key observable in the spin-0 CP analysis of the Higgs
resonance~\cite{wbf_spin}. On the other hand, the same is true for the
rapidity difference $\Delta y_{12}$ when it comes to spin-2
alternatives~\cite{wbf_spin}.

\section{Replacing a jet veto}
\label{sec:veto}

The key physics question we will answer in this Section is to what
degree we can use information on additional (central) jet radiation to
enhance the tagging jet analysis described in the previous
Section~\ref{sec:2jets}. Because a detailed analysis of the jet
activity has not been performed in the recent LHC runs, we assume a
collider energy of 13~TeV in this section.  The physics of the
additional jets can be easily described: for the signal events the
emission of additional central jets is suppressed by the color
structure of the process. This means that the number of jets in weak
boson fusion will in general follow the staircase pattern predicted
for inclusive processes at the LHC~\cite{scaling}. In contrast,
gluon--fusion Higgs production or di-photon production will show this
staircase pattern only in the absence of tagging jet cuts. Once we
require two hard jets with a large invariant mass we
induce large logarithms, which leads to a Poisson pattern in the number
of jets~\cite{scaling}. The key feature of this Poisson distribution
is a significantly enhanced probability of radiating a central
jet.\bigskip

Throughout our analysis we require two tagging jets with the generic
acceptance cuts
\begin{alignat}{5}
p_{Tj} &> 20~\gev
\qqqquad 
&|y_j| &< 4.5 
\label{eq:general_cuts_a}
\\
|\Delta y_{j_1j_2}| &> 2  
\qqqquad  
&m_{j_1j_2} &> 150~\gev
\; . 
\label{eq:general_cuts_b}
\end{alignat}
Correspondingly, we generate signal and background events using
\textsc{Sherpa}~\cite{sherpa} with \textsc{CKKW}~\cite{ckkw} jet
merging with two or three hard jets from the matrix element.
Throughout this Section we assume a collider energy of 13~TeV. In
addition to the general photon cuts of Eq.\eqref{eq:photon_cuts} we
require $m_{\gamma\gamma} = 126 \pm 10$~GeV. The cuts of
Eq.~\eqref{eq:general_cuts_a} lead to cross sections of 6.5~fb for the
weak--boson--fusion signal, 4.5~fb for gluon--fusion Higgs production,
and 2050~fb for the continuum background.  As mentioned above, the
signal--to--background ratio can be improved through additional cuts,
such as tightening the $m_{\gamma \gamma}$ requirement. However, this
makes it harder to reliably simulate the background. In the
following we will assume that additional cuts on the Higgs decay
products are orthogonal to the additional jet kinematics.

Because the selection criterion of the two tagging jets has a
significant impact on the amount of Poisson enhancement of the
additional jet production we use two selection criteria for the
tagging jets:
\begin{enumerate}
\item $p_T$-selection: of all jets fulfilling
  Eqs.\eqref{eq:general_cuts_a} and~\eqref{eq:general_cuts_b} the two
  hardest are the tagging jets.  The mild cuts of
  Eq.\eqref{eq:general_cuts_b} leave 3.36~fb for the signal, 1.04~fb
  for gluon--fusion Higgs production, and 509~fb for the continuum
  background.
\item $\Delta y$-selection: of all jets fulfilling
  Eq.\eqref{eq:general_cuts_a} and~\eqref{eq:general_cuts_b} the two
  most forward and backward are the tagging jets, maximizing $\Delta y_{j_1 j_2}$.
  After Eq.\eqref{eq:general_cuts_b} the remaining rates are 3.78~fb
  for the signal, 1.71~fb for gluon--fusion Higgs production, and
  736.2~fb for the non-Higgs background.
\end{enumerate}
While the $p_T$-selection is standard in most weak--boson--fusion
analyses, it will turn out that the $\Delta y$-selection is more
efficient in generating a large Poisson enhancement for central jet
emission in the background processes. On the other hand, in particular
for the 13~TeV run we have to see if pile-up makes one of the two
selections appear experimentally superior.\bigskip

\begin{table}[t]
\centering
\begin{tabular}{l|rrr|rrr}
\hline
& \multicolumn{3}{c|}{$\Delta y$-selection} & \multicolumn{3}{c}{$p_T$-selection} \\ 
\hline
& WBF & GF & $\gamma\gamma$ & WBF & GF & $\gamma\gamma$\\
\hline
generated [fb] & 6.5 & 4.5 & 2050 & 6.5 & 4.5 & 2050 \\ 
$\Delta y_{j_1j_2} > 4.4$   & $\times 0.33$ & $\times 0.15$ & $\times 0.11$ & $\times 0.27$ & $\times 0.056$ &  $\times 0.055$ \\
$y_{j1} y_{j2} < 0.0 $   & $\times 1.00$ & $\times 1.00$ & $\times 1.00$ & $\times 1.00$ & $\times 1.00$ &  $\times 1.00$ \\
$ m_{j_1j_2} > 600$~GeV& $\times 0.72$ & $\times 0.55$ & $\times 0.46$ & $\times 0.77$ & $\times 0.61$ &  $\times 0.47$ \\
cut level [fb] & 1.52 & 0.37 & 107 & 1.36 & 0.15 & 52.9 \\
central jet veto & $\times 0.75$ & $\times 0.15$ & $\times 0.22$ & $\times 0.91$ & $\times 0.45$ &  $\times 0.52$ \\ \hline
veto level [fb] &1.14 & 0.056 &24.0 &1.24 & 0.068 & 27.7\\ \hline 
\end{tabular}
\caption{Cut flow for the standard weak--boson--fusion analysis with a
  central jet veto for an LHC energy of 13~TeV.}
\label{tab:classical_veto}
\end{table}

\begin{table}[t]
\begin{tabular} {l|c|c|c|c||c|c|c|c}
\hline
& \multicolumn{4}{c||}{$\Delta y$-selection} 
& \multicolumn{4}{c}{$p_T$-selection} \\ \hline
& $\epsilon_S$ & $1 - \epsilon_B$ & $\dfrac{S}{\sqrt{S+B}}$ & $\dfrac{S}{B}$ 
& $\epsilon_S$ & $1 - \epsilon_B$ & $\dfrac{S}{\sqrt{S+B}}$ & $\dfrac{S}{B}$ \\ 
\hline
acceptance cuts Eqs.\eqref{eq:general_cuts_a} and~\eqref{eq:general_cuts_b}
& 1 & 0 & 0.76 & 0.005
& 1 & 0 & 0.81 & 0.007 \\
veto--level cuts Eq.\eqref{eq:veto_cuts} 
& 0.402  & 0.854  & 0.80 & 0.014 
& 0.405  & 0.996  & 1.01 & 0.026 \\ 
jet veto 
& 0.302 & 0.967 & 1.24 & 0.047 
& 0.369 & 0.945 & 1.26 & 0.045 \\\hline
\multirow{2}{5cm}{BDT: WBF default with Eq.\eqref{eq:general_cuts_b}} 
& 0.400  & 0.862 & 0.79 & 0.014  
& 0.400  & 0.904 & 1.04 & 0.027  \\ 
& 0.634  & 0.674 & 0.84 & 0.010  
& 0.414  & 0.897 & 1.04 & 0.027  \\ \hline
\multirow{2}{6.5cm}{BDT: WBF default plus FWM with Eq.\eqref{eq:general_cuts_b}} 
& 0.400  & 0.952 & 1.34 & 0.041 
& 0.400  & 0.944 & 1.35 & 0.047 \\ 
& 0.232  & 0.986 & 1.42 & 0.083
& 0.302  & 0.972 & 1.43 & 0.071 \\ \hline
\end{tabular}
\caption{$S/B$ and $S/\sqrt{S+B}$ compared to classical cut and jet
  veto strategy for the $\Delta y$ and $p_T$-selections of the tagging
  jets. The value for $S/\sqrt{S+B}$ we compute for an integrated
  luminosity of $30~\ifb$. The BDT analysis includes a set of
  Fox--Wolfram moments with unit weight,
  Eq.\eqref{eq:def_fwm_veto}. We quote two working points at 40\%
  signal efficiency and optimized for $S/\sqrt{S+B}$.}
\label{tab:veto_results}
\end{table}

The standard approach to including the additional jet activity in the
weak--boson--fusion Higgs analysis is a central jet veto~\cite{tagging,jetveto}.
To generate a sufficiently strong Poisson pattern in the number of
jets we demand
\begin{alignat}{5}
|\Delta y_{j_1j_2}| > 4.4 
\qqqquad 
y_{j1} \cdot y_{j2} < 0
\qqqquad
m_{j_1j_2} > 600~\text{GeV} \; .
\label{eq:veto_cuts}
\end{alignat}
In Table~\ref{tab:classical_veto} we show the cut flow of the signal
and background rates for each step in Eq.\eqref{eq:veto_cuts}.
Finally, we include a central jet veto which does not allow for jets
above $p_T=20$~GeV in between the two tagging jets. While the two
tagging jet selections show significant differences in the
intermediate steps, after the veto the numbers of signal and
background events are comparable. The survival rates for the central
jet veto are in agreement with the
literature~\cite{jetveto,wbf_tau}.\bigskip

In the first three rows of Table~\ref{tab:veto_results} we show different
statistical measures after the acceptance cuts of
Eqs.\eqref{eq:general_cuts_a} and~\eqref{eq:general_cuts_b}, the
veto--level cuts of Eq.\eqref{eq:veto_cuts}, and after the central jet
veto. The background is composed of gluon--fusion Higgs production and
continuum di-photon production. We again see that the significance
$S/\sqrt{S+B}$ and the signal--to--background ratio are comparable for
the $\Delta y$-selection and the $p_T$-selection of the tagging
jets. However, this is only true after the jet veto. After only the
hard cuts of Eq.\eqref{eq:veto_cuts} the $p_T$-selection is
significantly more promising. As alluded to above, the jet veto
benefits from the stronger Poisson enhancement from the $\Delta
y$-selection, leaving the final results essentially identical.

\begin{figure}[t]
\centering
\includegraphics[width=0.34\textwidth]{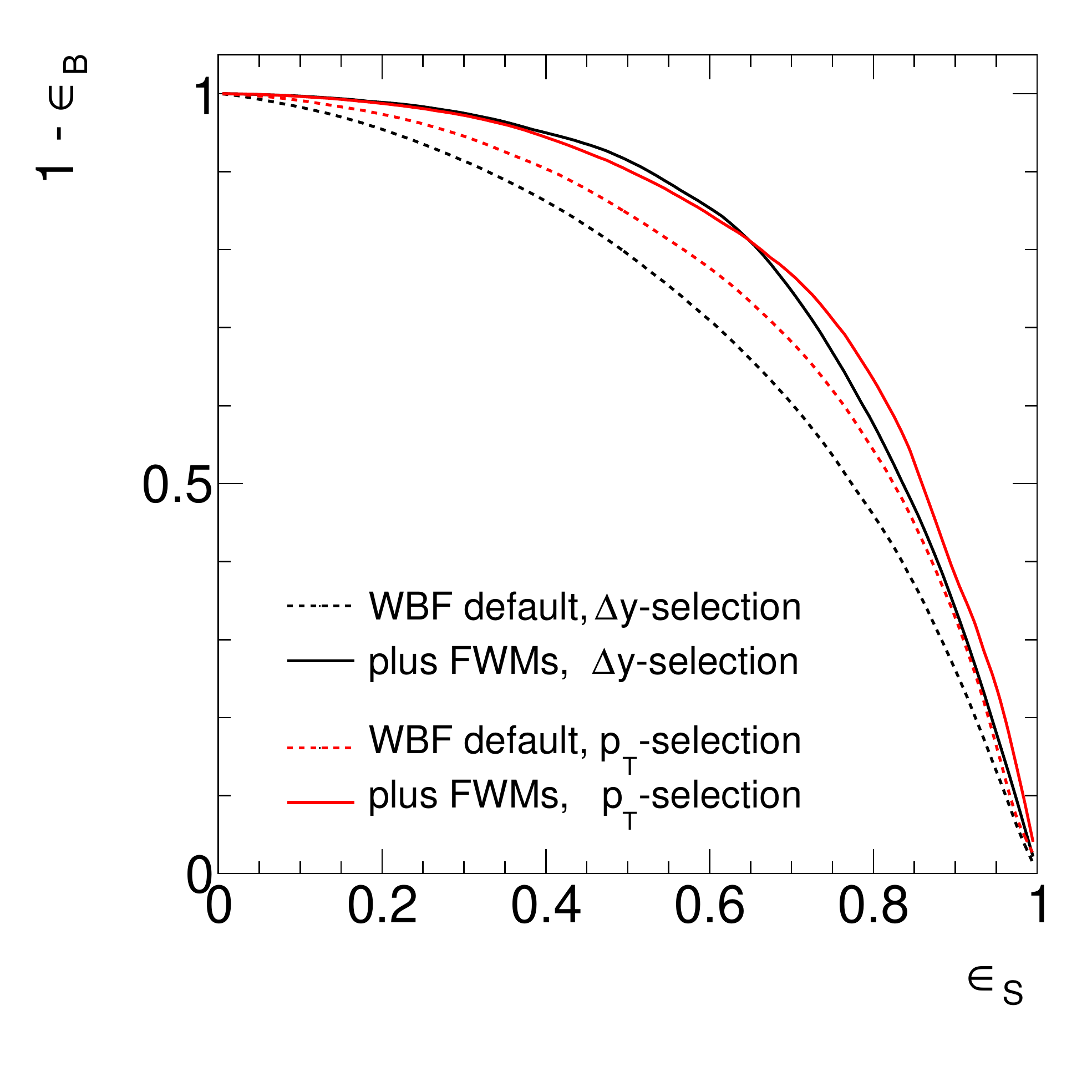}
\hspace*{0.1\textwidth}
\includegraphics[width=0.34\textwidth]{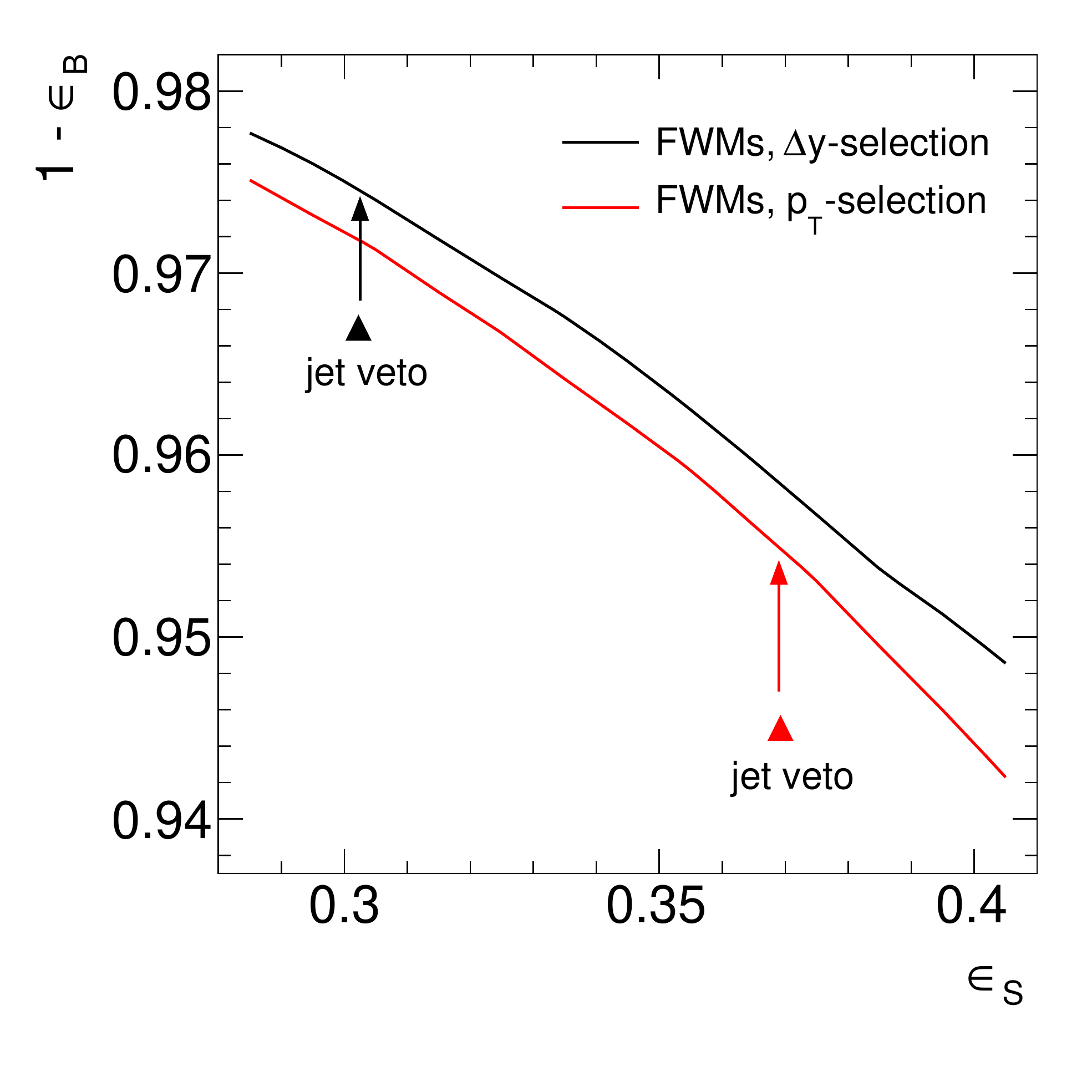}
\vspace*{-6mm}
\caption{ROC curve for $\Delta y$- (black) and $p_T$-selection (red)
  of the tagging jets. Left: We compare the WBF default observables
  (dashed) of Eq.\eqref{eq:wbf_default} to an additional set of
  Fox--Wolfram moments (solid). Right: We show how using Fox--Wolfram
  moments compare to a central jet veto.}
\label{fig:veto_roc}
\end{figure}

In the next step, we use the default WBF observables of
Eq.\eqref{eq:wbf_default} and optimize them in a multivariate BDT
analysis as described in Section~\ref{sec:bdt}. The corresponding ROC
curve we show in Figure~\ref{fig:veto_roc}. As in
Table~\ref{tab:veto_results} the efficiencies are defined with respect
to the full set of acceptance cuts from Eqs.\eqref{eq:general_cuts_a}
and~\eqref{eq:general_cuts_b}. In the table we quote two points from
this curve. First, we show the usual working point with a signal
efficiency of 40\%. Second, we show the working point with the best
result for $S/\sqrt{S+B}$. Optimizing for the best result of $S/B$
does not give a well defined solution. As expected, the ROC curve
indicates working points for the entire range of signal efficiencies 
$\epsilon_S = 0...1$.\bigskip

The question we attempt to answer in this section is if we can use the
available information on the additional jet activity in reducing the
two backgrounds more efficiently than with a central jet veto. The
baseline for this comparison is the corresponding row in
Table~\ref{tab:veto_results}. As described in
Section~\ref{sec:moments} we rely on a large set of Fox--Wolfram
moments forming a basis for the angular correlations given a weight
$W_{ij}^x$. Unlike for the tagging jet kinematics we now do not
constrain our system to the transverse plane, which means we use the
original definition of the moments in Eq.\eqref{eq:fwm_def2} with the
opening angle $\Omega_{ij}$. On the other hand, we already know what
the benefit of including the moments of the tagging jets are:
according to Section~\ref{sec:2jets} most of the information is
included once we add the azimuthal angle between the tagging jets,
$\Delta \phi_{j_1 j_2}$, to the standard set of observables given in
Eq.\eqref{eq:wbf_default}. Therefore, we limit the analysis of the
additional jet activity to all jet--jet correlations {\sl with the
  exception of the two tagging jets.} Moreover, we can expect the unit
weight to give the best sensitivity to the relatively soft additional
jet activity, so we use
\begin{alignat}{5}
  H^U_\ell &= \frac{1}{N^2} \; \sum \limits_{(i,j) \ne (1,2)} 
  P_\ell( \cos \Omega_{ij} ) \; .
\label{eq:def_fwm_veto}
\end{alignat}
For both of the tagging jet selections we only include jets which fall
between the two tagging jets, in complete analogy of a central jet
veto.  For exactly two tagging jets and no additional jet radiation
this implies $H^U_\ell = 0$ for all values of $\ell$.\bigskip

In Table~\ref{tab:veto_results} we show the result of a combined BDT
analysis of the observable of Eq.\eqref{eq:wbf_default} and the set of
Fox--Wolfram moments. Again, we quote two operating points, one of them
for a fixed signal efficiency of 40\% and one optimized for the best
value of $S/\sqrt{S+B}$. In addition, we show results for both, the
$\Delta y$-selection and the $p_T$-selection of the tagging jets. A
generic problem for any BDT analysis is that for limited statistics
of the training sample it can only include a limited number of
observables. On the other hand, the BDT first determines the most
powerful observables, so we only include the five best Fox--Wolfram
moments in our analysis. We have checked that adding more moments will
not improve the result beyond numerical accuracy. For the $\Delta
y$-selection the five leading moments with unit weight are $H^U_2$,
$H^U_4$, $H^U_{18}$, $H^U_{19}$, and $H^U_{17}$.  For the
$p_T$-selection the most powerful moments are $H^U_2$, $H^U_{19}$,
$H^U_{17}$, $H^U_{20}$, and $H^U_{15}$. However, for the
$p_T$-selection the most powerful variable in the BDT is $\Delta
y_{j_1 j_2}$. For the $\Delta y$-selection this observable is
maximized by construction.\bigskip

The ROC curves in Figure~\ref{fig:veto_roc} shows a clear improvement
of the complete multivariate analysis including the Fox--Wolfram
moments as compared to the kinematic variables of
Eq.\eqref{eq:wbf_default} only. For a fixed moderate signal efficiency
of 40\% adding information on the jets decreases the probability of a
background mis-identification by a factor of 2.9 for the $\Delta
y$-selection and a factor of 1.7 for the $p_T$-selection.  The
improvement relative to the jet veto we show in the right panel,
zooming into typical signal efficiencies around 35\% relative to the
acceptance cuts of Eq.\eqref{eq:general_cuts_b}. For the jet veto
working point of the $\Delta y$-selection with fixed signal efficiency
of 30.2\% we see that the background misidentification is reduced by
30\%. For the $p_T$-selection with fixed signal efficiency of 36.9\%
we find an improvement by 20\%.

\section{Avoiding new scales}
\label{sec:matched}

The unit weights in the definition of the Fox--Wolfram moments used in
the previous Section~\ref{sec:veto} share a disadvantage with a jet
veto when it comes to predicting them from theory: they introduce an
additional physical momentum scale in the process which is below the
hard scale of the Higgs production process. Collinear factorization as
the basis of defining the parton densities in perturbative field
theory does not allow for such additional scales. All measurements
which are to be compared to fixed--order perturbative QCD predictions
have to be jet--inclusive for transverse momenta below the
factorization scale. If we introduce an additional scale this implies
that we introduce a possibly large logarithm which needs to be
resummed~\cite{neubert,gavin,scet}.\bigskip

\begin{table}[b!]
\begin{tabular} {l|c|c|c|c||c|c|c|c}
\hline
& \multicolumn{4}{c||}{$\Delta y$-selection} 
& \multicolumn{4}{c}{$p_T$-selection} \\ \hline
& $\epsilon_S$ & $1 - \epsilon_B$ & $\dfrac{S}{\sqrt{S+B}}$ & $\dfrac{S}{B}$ 
& $\epsilon_S$ & $1 - \epsilon_B$ & $\dfrac{S}{\sqrt{S+B}}$ & $\dfrac{S}{B}$ \\ 
\hline
jet veto Eq.\eqref{eq:general_cuts_a} to \eqref{eq:veto_cuts}
& 0.302 & 0.967 & 1.24 & 0.047 
& 0.369 & 0.945 & 1.26 & 0.045 \\\hline
\multirow{2}{6.9cm}{BDT: WBF default plus unit--weight FWM} 
& 0.400  & 0.952 & 1.34 & 0.041 
& 0.400  & 0.944 & 1.35 & 0.047 \\ 
& 0.232  & 0.986 & 1.42 & 0.083
& 0.302  & 0.972 & 1.43 & 0.071 \\\hline
\multirow{2}{6.9cm}{BDT: WBF default plus matched--weight FWM} 
& 0.400  & 0.949 & 1.32 & 0.040  
& 0.400  & 0.942 & 1.32 & 0.045  \\ 
& 0.240  & 0.985 & 1.43 & 0.081 
& 0.256  & 0.979 & 1.40 & 0.082  \\\hline
\end{tabular}
\caption{$S/B$ and $S/\sqrt{S+B}$ compared to jet veto strategy for
  the $\Delta y$ and $p_T$-selections of the tagging jets. The value
  for $S/\sqrt{S+B}$ we compute for an integrated luminosity of
  $30~\ifb$. Extending Table~\ref{tab:veto_results} the BDT analysis
  now includes a set of Fox--Wolfram moments with matched weight,
  Eq.\eqref{eq:def_fwm_matched}.  As BDT results we quote the working
  point at 40\% signal efficiency and the best point for
  $S/\sqrt{S+B}$.}
\label{tab:matched_results}
\end{table}

Introducing a weight which smoothly interpolates between the jet counting 
scale $p_{T j}^\text{min}=20$~GeV and the hard scale of the process
according to Eq.\eqref{eq:weight_2} should alleviate this tension,
suggesting to repeat the same analysis as shown in
Section~\ref{sec:veto} with the Fox--Wolfram moments
\begin{alignat}{1}
H^M_\ell &= \sum \limits_{(i,j) \ne (1,2)} 
\frac{(p_{T i}-p_T^\text{min}) \, (p_{T j}-p_T^\text{min})}
     {\left(\sum p_{T i} - p_T^\text{min} \right)^2} \; 
P_\ell( \cos \Omega_{ij} ) \; .
\label{eq:def_fwm_matched}
\end{alignat}
While we cannot offer an estimate of the improvement in the
perturbative QCD treatment, it is clear that the matched weights are
less sensitive to large collinear logarithms generated by the
violation of collinear factorization.

\begin{figure}[t]
\centering \includegraphics[width=0.34\textwidth]{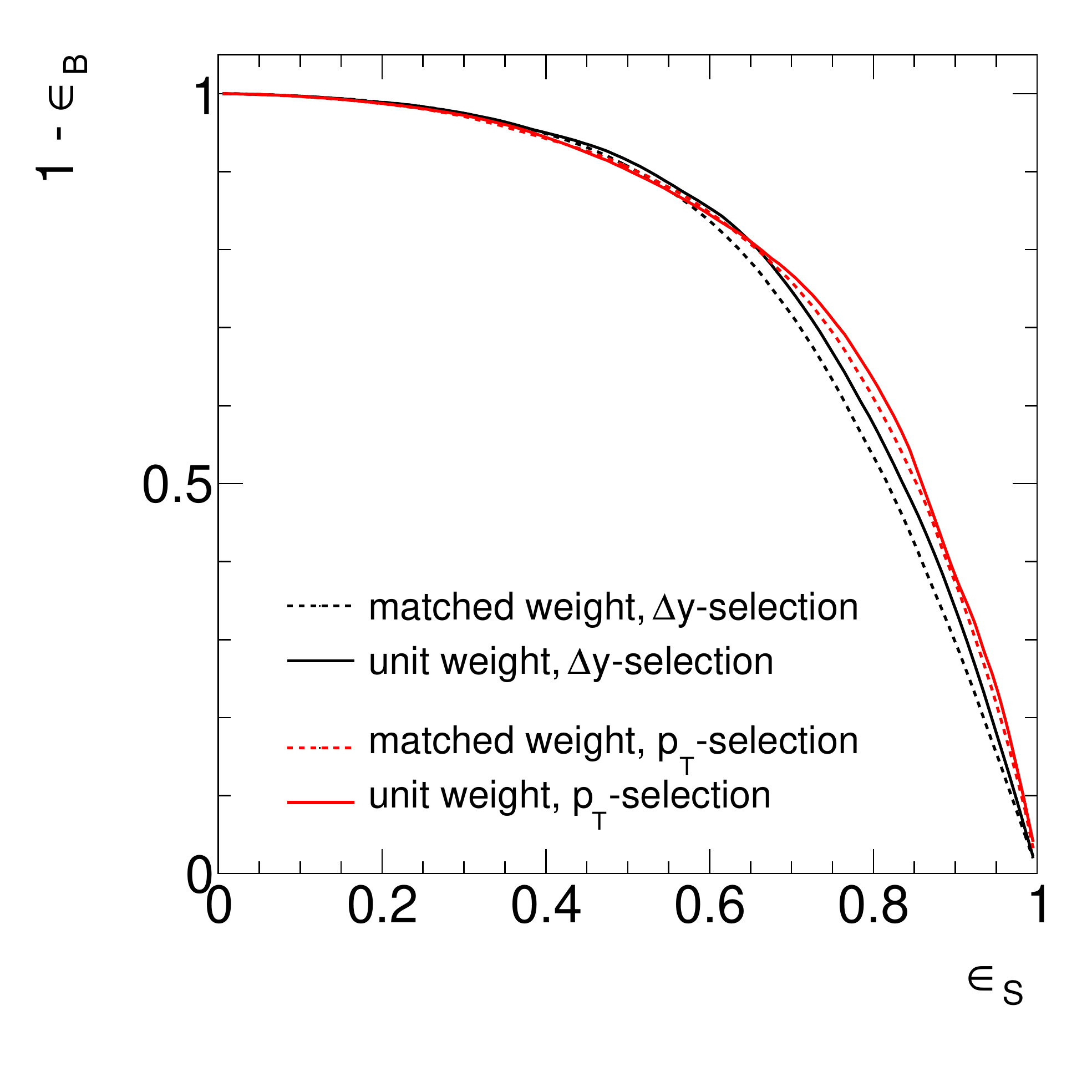}
\hspace*{0.1\textwidth}
\includegraphics[width=0.34\textwidth]{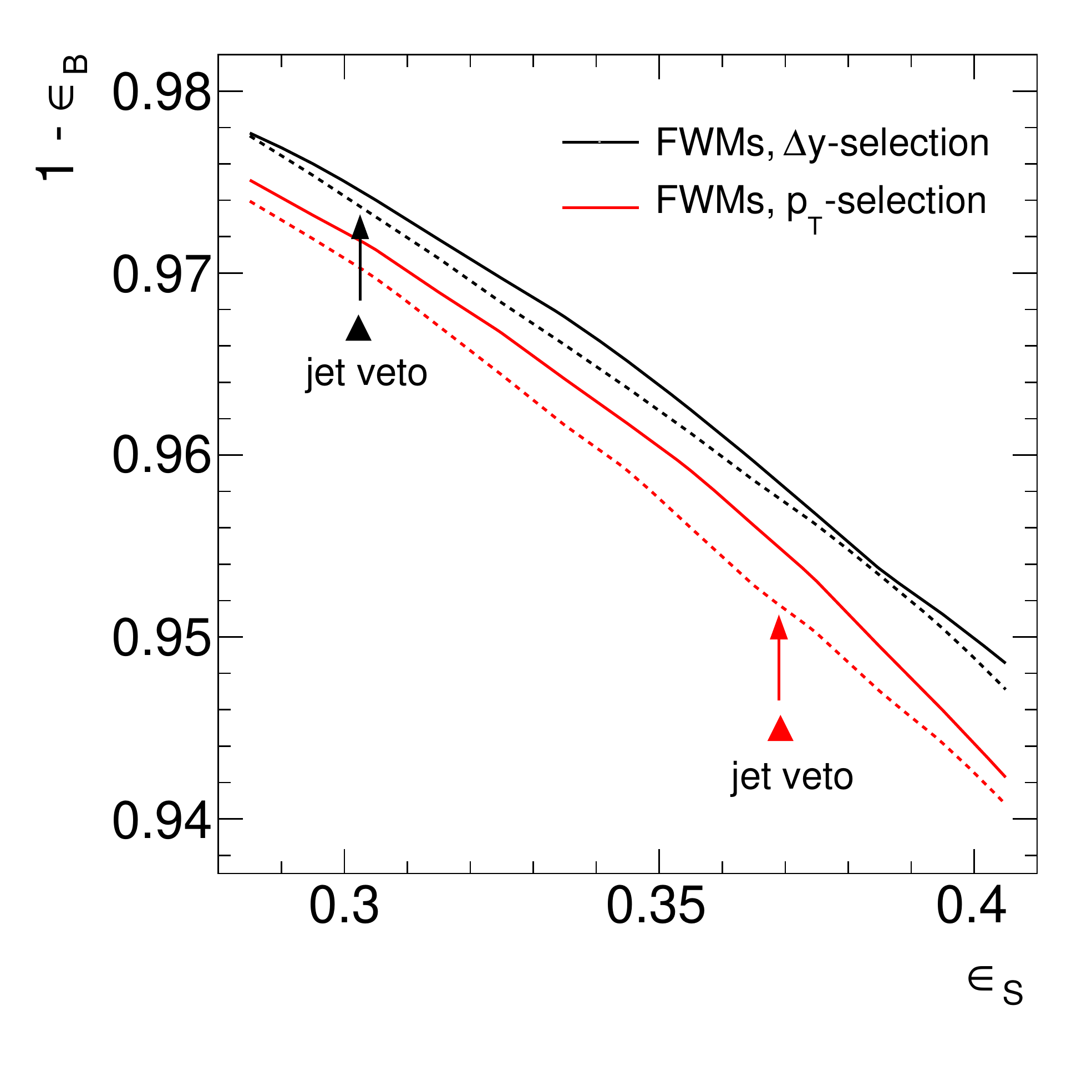}
\vspace*{-6mm}
\caption{ROC curve for $\Delta y$- (black) and $p_T$-selection (red)
  of the tagging jets. Left: We compare the WBF default observables
  (dashed) of Eq.\eqref{eq:wbf_default} to an additional set of
  Fox--Wolfram moments (solid). Right: We show how using Fox--Wolfram
  moments compares to a central jet veto.}
\label{fig:matched_roc}
\end{figure}

In Table~\ref{tab:matched_results} we extend the original
Table~\ref{tab:veto_results}, including the same BDT analysis now
based on matched Fox--Wolfram moments. For the standard working point
with 40\% signal efficiency we see that the background rejection from
the matched moments is essentially identical to the unit weight
moments. The main difference is the order of the most relevant set of
moments, which now is $H^M_1,H^M_2,H^M_3,H^M_4,H^M_6$ for the $\Delta
y$-selection and $H^M_1,H^M_3,H^M_6,H^M_2,H^M_4$ for the
$p_T$-selection. Similarly, the working point optimized for
$S/\sqrt{S+B}$ is only slightly shifted. In
Figure~\ref{fig:matched_roc} we compare the ROC curves for the jet
radiation study based on the two Fox--Wolfram moment weights. For
signal efficiencies between 25\% and 40\% the unit weight is slightly
superior, but most likely this slight advantage will be compensated 
once we include theory uncertainties from QCD predictions.

\vspace{-1em}
\section{Outlook}

Weak boson fusion analyses of Higgs production at the LHC are key
ingredients to Higgs couplings and Higgs property analyses in the
upcoming LHC run. They allow for an efficient background rejection
based on two tagging jets and an additional central jet veto. The
question is, how we can make optimal use of the jet properties for
example to improve the signal--to--background ratio or the signal
significance. In our detailed analysis we come to three conclusions:
\begin{enumerate}
\item For the two tagging jets we rely on a set of low-$\ell$ moments
  with a transverse momentum weight and azimuthal angle
  separation. Most of the improvement as compared to the standard
  ATLAS analysis can be traced back to the missing azimuthal angle
  between the tagging jets. In addition, the signal--to--background
  ratio can be increased by 8\% by including a set of Fox--Wolfram
  moments.
\item The additional jets can be studied using a wide range of moments
  with a unit weight and full angular separation. It should be
  compared to a jet veto and delivers a significantly better
  performance. The tagging jet selection with maximum rapidity
  distance is better suited to distinguish the signal from the
  continuum background then the transverse momentum selection. For
  both cases we computed a full ROC curve, allowing for optimized
  working points depending on the details of the analysis.
\item To reduce theory uncertainties from QCD predictions we can
  introduce a softer, matched weight in the Fox--Wolfram moments. It
  turns out that the analysis of jet radiation is almost as promising
  as for the unit weights, but with a much improved theoretical
  behavior.
\end{enumerate}
We conclude that tagging jet criteria as well as the jet veto as
analysis tools for Higgs analyses in weak boson fusion can be improved
by a systematic study of the multi--jet system based on Fox--Wolfram
moments. The improvement is significant, both for the $\Delta
y$-selection and the $p_T$-selection of the tagging jets.  The
Fox--Wolfram moment analysis can be adapted to individual analyses by
choosing appropriate working points in the corresponding ROC
curves.\bigskip

\newpage

\begin{center}
{\bf Acknowledgments}
\end{center}

CB would like to thank the ATLAS collaboration at CERN, where part of
this work was performed. We would like to thank Erik Gerwick and
Steffen Schumann for help with the event generation. CB acknowledges
support by BMBF under project number 05H12VHE. PS acknowledges support by the
IMPRS for \emph{Precision Tests of Fundamental Symmetries}.



\begin{thebibliography}{99}

\bibitem{higgs}
 P.~W.~Higgs,
  Phys.\ Lett.\  {\bf 12}, 132 (1964);
 P.~W.~Higgs,
  Phys.\ Rev.\ Lett.\  {\bf 13}, 508 (1964);
 F.~Englert and R.~Brout,
  Phys.\ Rev.\ Lett.\  {\bf 13}, 321 (1964).

\bibitem{atlas}
 G.~Aad {\it et al.}  [ATLAS Collaboration],
  Phys.\ Lett.\ B  {\bf{716}}, 1 (2012).

\bibitem{cms}
 S.~Chatrchyan {\it et al.}  [CMS Collaboration],
  Phys.\ Lett.\ B  {\bf{716}}, 30 (2012).

\bibitem{ilc}
 G.~Weiglein {\it et al.}  [LHC/LC Study Group Collaboration],
  Phys.\ Rept.\  {\bf 426}, 47 (2006);
 M.~Klute, R.~Lafaye, T.~Plehn, M.~Rauch and D.~Zerwas,
  Europhys.\ Lett.\  {\bf 101}, 51001 (2013).

\bibitem{rge_studies}
 see \eg 
 M.~Shaposhnikov and C.~Wetterich,
  Phys.\ Lett.\ B {\bf 683}, 196 (2010);
 M.~Holthausen, K.~S.~Lim and M.~Lindner,
  JHEP {\bf 1202}, 037 (2012);
 A.~Hebecker, A.~K.~Knochel and T.~Weigand,
  Nucl.\ Phys.\ B {\bf 874}, 1 (2013)
 D.~Buttazzo {\it et al}
  arXiv:1307.3536 [hep-ph];

\bibitem{couplings}
 for up-to-date coupling analyses see 
 [ATLAS Collaboration],
 ATLAS-CONF-2013-034;
 [ATLAS Collaboration],
 ATLAS-CONF-2013-040;
 [CMS Collaboration],	
 CMS-PAS-HIG-13-005;
 T.~Plehn and M.~Rauch,
  Europhys.\ Lett.\  {\bf 100}, 11002 (2012);
 D.~Lopez-Val, T.~Plehn and M.~Rauch,
  JHEP {\bf 1310}, 134 (2013);
 T.~Corbett, O.~J.~P.~Eboli, J.~Gonzalez-Fraile and M.~C.~Gonzalez-Garcia,
  arXiv:1306.0006 [hep-ph];
 I.~Brivio {\it et al}
  arXiv:1311.1823 [hep-ph].
 A.~Azatov, R.~Contino and J.~Galloway,
  JHEP {\bf 1204}, 127 (2012);
 J.~Ellis and T.~You,
  JHEP {\bf 1306}, 103 (2013);
 G.~F.~Giudice, C.~Grojean, A.~Pomarol and R.~Rattazzi,
  JHEP {\bf 0706}, 045 (2007);
 J.~R.~Espinosa, C.~Grojean, M.~M\"uhlleitner and M.~Trott,
  JHEP {\bf 1212}, 045 (2012);
 S.~Banerjee, S.~Mukhopadhyay and B.~Mukhopadhyaya,
  JHEP {\bf 1210}, 062 (2012);
 N.~Craig and S.~Thomas,
  JHEP {\bf 1211}, 083 (2012);
 F.~Bonnet, T.~Ota, M.~Rauch and W.~Winter,
  Phys.\ Rev.\ D {\bf 86}, 093014 (2012);
 A.~Djouadi,
  arXiv:1208.3436 [hep-ph];
 B.~A.~Dobrescu and J.~D.~Lykken,
  JHEP {\bf 1302}, 073 (2013);
 E.~Mass\'o and V.~Sanz,
  Phys.\ Rev.\ D {\bf 87}, 033001 (2013);
 G.~Belanger, B.~Dumont, U.~Ellwanger, J.~F.~Gunion and S.~Kraml,
  JHEP {\bf 1302}, 053 (2013);
 P.~P.~Giardino, K.~Kannike, I.~Masina, M.~Raidal and A.~Strumia,
  arXiv:1303.3570 [hep-ph]; 
 A.~Djouadi and G.~Moreau,
  arXiv:1303.6591 [hep-ph];
 D.~Carmi, A.~Falkowski, E.~Kuflik, T.~Volansky and J.~Zupan,
  JHEP {\bf 1210}, 196 (2012);
 P.~P.~Giardino, K.~Kannike, I.~Masina, M.~Raidal and A.~Strumia,
  arXiv:1303.3570 [hep-ph]; 

\bibitem{wbf_tau}
 D.~L.~Rainwater, D.~Zeppenfeld and K.~Hagiwara,
  Phys.\ Rev.\  D {\bf 59}, 014037 (1999);
 T.~Plehn, D.~L.~Rainwater and D.~Zeppenfeld,
  Phys.\ Rev.\  D {\bf 61}, 093005 (2000).

\bibitem{wbf_w}
 N.~Kauer, T.~Plehn, D.~Rainwater and D.~Zeppenfeld,
  Phys.\ Lett.\ B {\bf 503}, 113 (2001).

\bibitem{wbf_gamma}
 D.~L.~Rainwater and D.~Zeppenfeld,
  JHEP {\bf 9712}, 005 (1997);
 J.~R.~Andersen, C.~Englert and M.~Spannowsky,
  arXiv:1211.3011 [hep-ph].

\bibitem{wbf_ex}
 S.~Asai, G.~Azuelos, C.~Buttar, V.~Cavasinni, D.~Costanzo, K.~Cranmer, R.~Harper and K.~Jakobs {\it et al.},
  Eur.\ Phys.\ J.\ C {\bf 32S2}, 19 (2004).

\bibitem{review}
 for LHC reviews \eg
 A.~Djouadi,
  Phys.\ Rept.\  {\bf 457}, 1 (2008);
 T.~Plehn,
  Lect.\ Notes Phys.\  {\bf 844}, 1 (2012)
  [arXiv:0910.4182 [hep-ph]].

\bibitem{tagging}
 R.~Kleiss and W.~J.~Stirling,
  Phys.\ Lett.\ B {\bf 200}, 193 (1988);
 U.~Baur and E.~W.~N.~Glover,
  Phys.\ Lett.\ B {\bf 252}, 683 (1990);
 V.~D.~Barger, K.~Cheung, T.~Han, J.~Ohnemus and D.~Zeppenfeld,
  Phys.\ Rev.\ D {\bf 44}, 1426 (1991).

\bibitem{scaling}
 E.~Gerwick, T.~Plehn and S.~Schumann,
  Phys.\ Rev.\ Lett.\  {\bf 108}, 032003 (2012);
 E.~Gerwick, T.~Plehn, S.~Schumann and P.~Schichtel,
  JHEP {\bf 1210}, 162 (2012);
 E.~Gerwick, S.~Schumann, B.~Gripaios and B.~Webber,
  JHEP {\bf 1304}, 089 (2013).

\bibitem{manchester}
 B.~E.~Cox, J.~R.~Forshaw and A.~D.~Pilkington,
  Phys.\ Lett.\ B {\bf 696}, 87 (2011).

\bibitem{jetveto}
 D.~L.~Rainwater, R.~Szalapski and D.~Zeppenfeld,
  Phys.\ Rev.\ D {\bf 54}, 6680 (1996).

\bibitem{neubert}
 T.~Becher and M.~Neubert,
  JHEP {\bf 1207}, 108 (2012);
 T.~Becher, M.~Neubert and L.~Rothen,
  JHEP {\bf 1310}, 125 (2013).

\bibitem{gavin}
 A.~Banfi, G.~P.~Salam and G.~Zanderighi,
  JHEP {\bf 1206}, 159 (2012)
 A.~Banfi, P.~F.~Monni, G.~P.~Salam and G.~Zanderighi,
  Phys.\ Rev.\ Lett.\  {\bf 109}, 202001 (2012).

\bibitem{scet}
 I.~W.~Stewart and F.~J.~Tackmann,
  Phys.\ Rev.\ D {\bf 85}, 034011 (2012);
 I.~W.~Stewart, F.~J.~Tackmann, J.~R.~Walsh and S.~Zuberi,
  arXiv:1307.1808 [hep-ph].

\bibitem{fwm_orig}
 G.~C.~Fox and S.~Wolfram,
  Phys.\ Rev.\ Lett.\  {\bf 41}, 1581 (1978);
 R.~D.~Field, Y.~Kanev and M.~Tayebnejad,
  Phys.\ Rev.\ D {\bf 55}, 5685 (1997).

\bibitem{first}
 C.~Bernaciak, M.~S.~A.~Buschmann, A.~Butter and T.~Plehn,
  Phys.\ Rev.\ D {\bf 87}, 073014 (2013).

\bibitem{wbf_spin}
 T.~Plehn, D.~L.~Rainwater and D.~Zeppenfeld,
  Phys.\ Rev.\ Lett.\  {\bf 88}, 051801 (2002);
 C.~Ruwiedel, N.~Wermes and M.~Schumacher,
  Eur.\ Phys.\ J.\ C {\bf 51}, 385 (2007);
 K.~Hagiwara, Q.~Li and K.~Mawatari,
  JHEP {\bf 0907}, 101 (2009);
 C.~Englert, D.~Goncalves-Netto, K.~Mawatari and T.~Plehn,
  JHEP {\bf 1301}, 148 (2013);
 A.~Djouadi, R.~M.~Godbole, B.~Mellado and K.~Mohan,
  Phys.\ Lett.\ B {\bf 723}, 307 (2013).


\bibitem{event_shapes}
 for a nice overview of LHC applications see \eg
 A.~Banfi, G.~P.~Salam and G.~Zanderighi,
  JHEP {\bf 1006}, 038 (2010).

\bibitem{tmva}
 A.~H\"ocker {\it et al.},
  PoS ACAT {\bf }, 040 (2007)
  [physics/0703039 [PHYSICS]];
 P.~Speckmayer, A.~H\"ocker, J.~Stelzer and H.~Voss,
  J.\ Phys.\ Conf.\ Ser.\  {\bf 219}, 032057 (2010);
 \url{http://tmva.sourceforge.net}

\bibitem{sherpa}
 T.~Gleisberg, S.~H\"oche, F.~Krauss, M.~Sch\"onherr, S.~Schumann, F.~Siegert and J.~Winter,
  JHEP {\bf 0902}, 007 (2009).

\bibitem{ckkw} 
 S.~Catani, F.~Krauss, R.~Kuhn and B.~R.~Webber,
  JHEP {\bf 0111}, 063 (2001).

\bibitem{fastjet}
 M.~Cacciari, G.~P.~Salam and G.~Soyez,
  Eur.\ Phys.\ J.\ C {\bf 72}, 1896 (2012).

\bibitem{atlas_default}
 [ATLAS Collaboration],
  ATLAS-CONF-2013-012.

\bibitem{C45}
 J.~R.~Quinlan,
 \emph{C4.5: Programs for Machine Learning},
 Morgan Kaufmann, San Mateo, CA (1992)

\end{thebibliography}
\end{document}